\documentclass[english]{article}
\usepackage{mathptmx}
\usepackage[T1]{fontenc}
\usepackage[latin9]{inputenc}
\usepackage{amsmath}
\usepackage{amsthm}
\usepackage{amssymb}
\usepackage{graphicx}
\usepackage[authoryear]{natbib}

\makeatletter
\theoremstyle{plain}
\newtheorem{thm}{\protect\theoremname}
\theoremstyle{definition}
\newtheorem{problem}[thm]{\protect\problemname}

\newcommand{\diag}{\mathop{\rm diag}}

\newcommand\trace{\mathop{\rm Tr}}

\renewcommand{\vec}[1]{{\bf #1}}

\makeatother

\usepackage{babel}
\providecommand{\problemname}{Problem}
\providecommand{\theoremname}{Theorem}

\begin{document}
\title{Two ``correlation games'' for a nonlinear network with Hebbian excitatory
neurons and anti-Hebbian inhibitory neurons}
\author{H. Sebastian Seung\\
Neuroscience Institute and Computer Science Dept.\\
Princeton University\\
Princeton, NJ 08544}
\maketitle
\begin{abstract}
A companion paper introduces a nonlinear network with Hebbian excitatory
($E$) neurons that are reciprocally coupled with anti-Hebbian inhibitory
($I$) neurons and also receive Hebbian feedforward excitation from
sensory ($S$) afferents. The present paper derives the network from
two normative principles that are mathematically equivalent but conceptually
different. The first principle formulates unsupervised learning as
a constrained optimization problem: maximization of $S-E$ correlations
subject to a copositivity constraint on $E-E$ correlations. A combination
of Legendre and Lagrangian duality yields a zero-sum continuous game
between excitatory and inhibitory connections that is solved by the
neural network. The second principle defines a zero-sum game between
$E$ and $I$ cells. $E$ cells want to maximize $S-E$ correlations
and minimize $E-I$ correlations, while $I$ cells want to maximize
$I-E$ correlations and minimize power. The conflict between $I$
and $E$ objectives effectively forces the $E$ cells to decorrelate
from each other, although only incompletely. Legendre duality yields
the neural network.
\end{abstract}
\begin{figure}
\quad{}\includegraphics[width=0.5\textwidth]{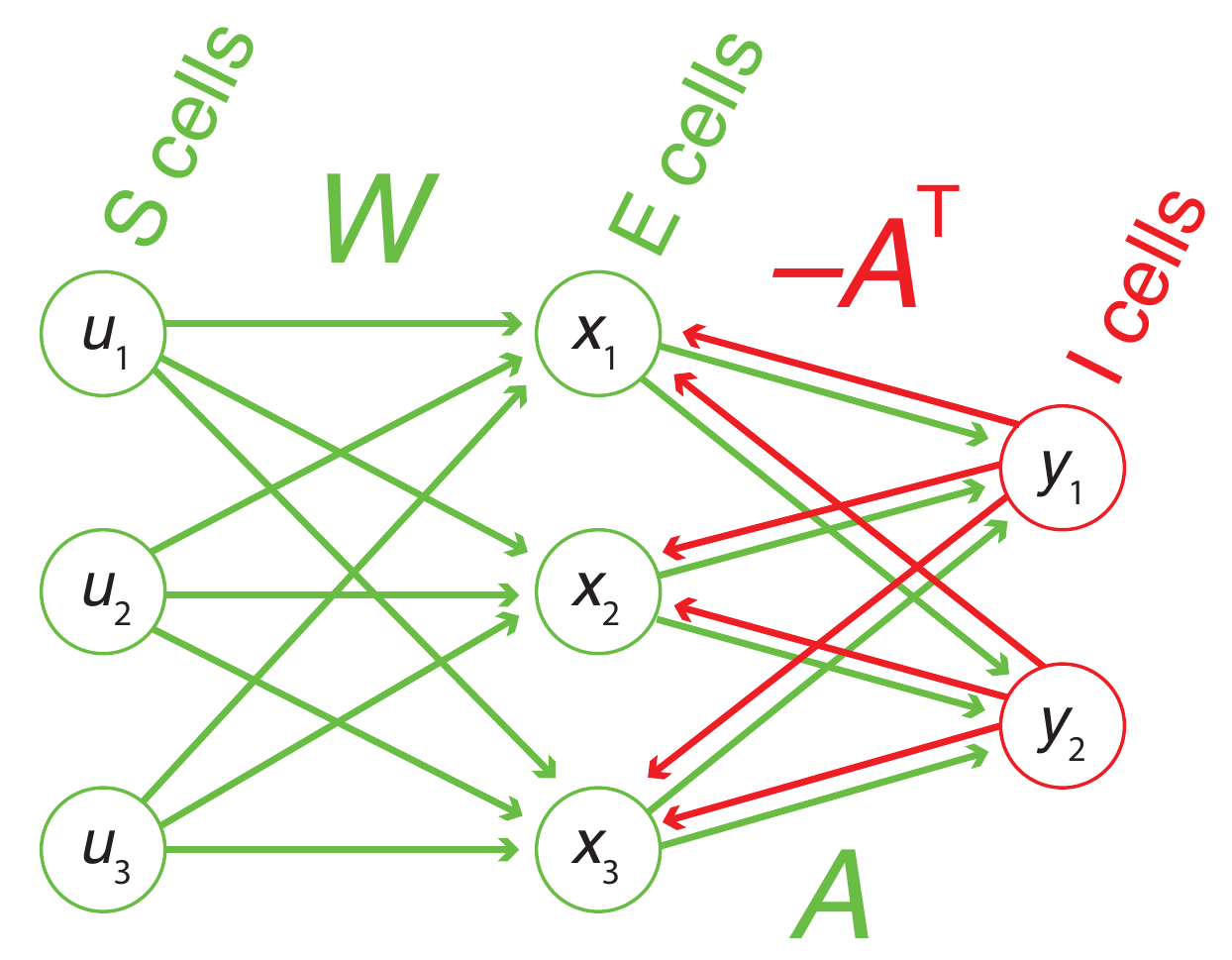}\qquad{}\includegraphics[width=0.5\textwidth]{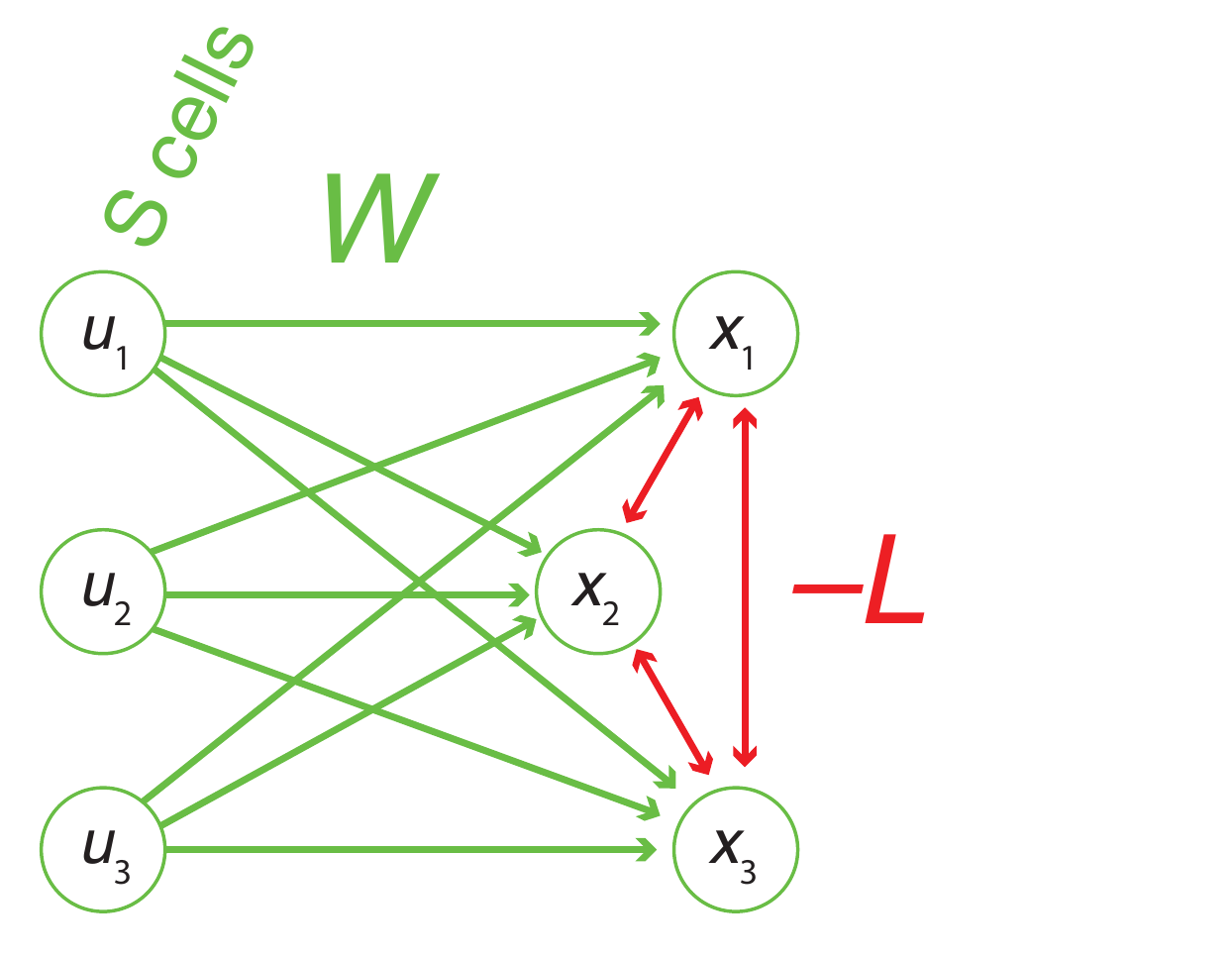}

\caption{Disynaptic inhibition versus direct all-to-all inhibition. Green arrows
indicate excitatory connections, which change via Hebbian plasticity.
Red arrows indicate inhibitory connections (note minus sign), which
change via anti-Hebbian plasticity. (Left) $E$ neurons are reciprocally
coupled with $I$ neurons, and receive input from $S$ afferents.
$E\to I$ connections ($A$) and $I\to E$ connections ($-A^{\top})$
have equal but opposite strengths. (Right) The $E$ and $I$ populations
collapse to a single population of neurons directly connected by all-to-all
inhibition ($-L$).\label{fig:NetworkArchitecture}}
\end{figure}

A companion paper \citep{seung2018unsupervised} introduces a nonlinear
neural network for unsupervised learning in which a population of
excitatory ($E$) neurons is reciprocally connected with a population
of inhibitory ($I$) neurons (Fig. \ref{fig:NetworkArchitecture},
left). The $E$ neurons also receive feedforward excitation from a
population of sensory ($S$) afferents. The reciprocal $E-I$ connections
allow the $E$ neurons to inhibit each other through disynaptic $E\to I\to E$
pathways mediated by $I$ neurons. This network motif will be called
``disynaptic recurrent inhibition,'' or just ``disynaptic inhibition.''
Excitatory connections ($S\to E$ and $E\to I$) are modified by Hebbian
plasticity, and inhibitory connections ($I\to E$) by anti-Hebbian
plasticity.

The companion paper investigates the computational properties of the
neural network through mathematical analysis and numerical simulations.
It would be helpful to attain a deeper understanding of the network
as arising from some normative principle of unsupervised learning.
As emphasized by \citet{pehlevan2015normative}, the long history
of neural network models based on Hebbian and anti-Hebbian plasticity
has primarily relied on numerical simulations to demonstrate interesting
self-organizing behaviors, and lacked normative principles that can
be mathematically formalized as optimizations of some objective function.
Normative principles are not only helpful for understanding, but also
have the practical consequence of suggesting optimization algorithms
other than neural networks, which could be useful in certain settings.

The companion paper takes one step towards a normative principle by
``deriving'' the network with disynaptic inhibition (Fig. \ref{fig:NetworkArchitecture},
left) as an approximation to a network with all-to-all inhibition
(Fig. \ref{fig:NetworkArchitecture}, right). The latter network contains
a single population of neurons that directly inhibit each other and
also receive feedforward excitation from sensory afferents, as in
the \citet{seung2017correlation} variant of the \citet{foldiak1990forming}
model. The derivation involves a factorized approximation of a Lagrange
multiplier matrix, which is not fully justified in the companion paper.
The present paper is intended to provide the missing theoretical foundation
by deriving the network from two normative principles that are mathematically
equivalent but suggest different interpretations.

The first normative principle is a modification of a principle previously
used by \citet{seung2017correlation} to derive the network with all-to-all
inhibition. They defined unsupervised learning as the maximization
of input-output correlations subject to a bound constraint on output-output
correlations. The bound constraint can be regarded as enforcing a
decorrelated output representation. The principle led to a zero-sum
``correlation game'' between excitation and all-to-all inhibition
in a nonlinear network that was a variant of the \citet{foldiak1990forming}
model.\footnote{\citet{zylberberg2011sparse} claimed that the \citet{foldiak1990forming}
model learns by fitting a linear generative model to sensory stimuli.
The claim is only an approximation at best, and invalid at worst.
It is conceptually very different from \citet{seung2017correlation}
and \citet{pehlevan2018similarity}, who view excitation and inhibition
as players in a zero-sum game or minimax problem. According to the
linear generative model idea, excitation and inhibition minimize reconstruction
error.} The inhibitory connections of the network were conjugate to output-output
correlations via Lagrangian duality. The excitatory connections were
conjugate to input-output correlations via Legendre duality. A similar
approach was taken by \citet{pehlevan2018similarity}, who derived
linear networks performing variants of principal component analysis
from the similarity matching principle.

For the disynaptic inhibition network, the first normative principle
is a constrained optimization problem: maximization of input-output
correlations subject to a copositivity constraint on output-output
correlations. The constraint can be regarded as forcing an imperfect
form of decorrelation. The principle leads to a zero-sum game solved
by the disynaptic inhibition network. The $S\to E$ connections are
conjugate to the input-output correlations via Legendre duality. The
$I-E$ connections are Lagrange multipliers enforcing the copositivity
constraint.

The first normative principle is based on the activities of the $E$
cells only; the activities of the $I$ cells do not appear until after
the duality transforms. The second normative principle is a zero-sum
game between $E$ and $I$ cells. $E$ cells want to maximize $S-E$
correlations and minimize $E-I$ correlations. $I$ cells want to
maximize $E-I$ correlations and minimize power. There is some conflict
inherent in the fact that $I$ and $E$ cells want to drive $E-I$
correlations in opposite directions. The compromise is that $E$ cells
tend to decorrelate from each other. Both excitatory and inhibitory
connections arise from Legendre duality.

A theatrical metaphor describes the difference between the normative
principles. The first normative principle places the $E$ cells center
stage while the $I$ cells lurk backstage. In the second normative
principle, $E$ and $I$ cells are both on stage; $E$ cells are leading
actors while $I$ cells are supporting actors. No connections appear
in either normative principle; the network only emerges after duality
transforms.

Both normative principles are mathematically equivalent to each other,
but they invite different viewpoints. The first normative principle
is simple because it contains $E$ cell activity only, but complex
because copositivity is not a terribly intuitive idea. The second
normative principle is complex because it contains both $E$ and $I$
cell activity, but simple because there is no need for copositivity.

The second normative principle is similar to two algorithms for dimensionality
reduction (hard-thresholding of covariance eigenvalues and equalizing
thresholded covariance eigenvalues) defined by \citet{pehlevan2015normative}.
The neural network implementation of the algorithms involved separate
populations of principal neurons and interneurons, but the neurons
did not obey Dale's Law and so could not be interpreted as excitatory
and inhibitory neurons.

\section{Network model with disynaptic inhibition\label{sec:NetworkModel}}

The disynaptic inhibition network (Fig. \ref{fig:NetworkArchitecture},
left) has the activity dynamics, 
\begin{align}
x_{i} & :=\left[\left(1-dt\right)x_{i}+dt\:\lambda_{i}^{-1}\left(\sum_{a=1}^{m}W_{ia}u_{a}-\sum_{\alpha=1}^{r}y_{\alpha}A_{\alpha i}\right)\right]^{+}\label{eq:ExcitatoryDynamics}\\
y_{\alpha} & =\sum_{i=1}^{n}A_{\alpha i}x_{i}\label{eq:InhibitoryActivity}
\end{align}
Here $dt$ is a step size parameter, which can be set at a small constant
value or adjusted adaptively. The activation function $\left[z\right]^{+}=\max\left\{ z,0\right\} $
is half-wave rectification. After the activities converge to a steady
state, update the connection matrices via

\begin{align}
\Delta W_{ia} & \propto x_{i}u_{a}-\gamma W_{ia}-\kappa\sum_{b}W_{ib}\label{eq:UpdateW}\\
\Delta A_{\alpha j} & \propto y_{\alpha}x_{j}-\left(q^{2}-p^{2}\right)A_{\alpha j}-p^{2}\sum_{i}A_{\alpha i}\label{eq:UpdateA}
\end{align}
where $\gamma>0$, $\kappa>0$, and $q^{2}>p^{2}$ . After the updates
(\ref{eq:UpdateW}) and (\ref{eq:UpdateA}), any negative elements
of $W$ and $A$ are zeroed to maintain nonnegativity. The divisive
factor $\lambda_{i}>0$ in Eq. (\ref{eq:ExcitatoryDynamics}) is updated
via

\begin{equation}
\Delta\lambda_{i}\propto x_{i}^{2}-q^{2}\label{eq:Homeostatic}
\end{equation}
Intuitions behind the model definitions are explained in the companion
paper \citep{seung2018unsupervised}. The goal of the present paper
is show how the network can be interpreted as a method of solving
a zero-sum game.

\section{Correlation game between connections}

\subsection{Formulation as constrained optimization}

The first normative principle concerns transformation of a sequence
of input vectors $\vec{u}(1),\ldots,\vec{u}(T)$ into a sequence of
output vectors $\vec{x}(1),\hdots,\vec{x}(T)$. Both input and output
are assumed nonnegative. Define the input matrix $U=\left[\vec{u}(1),\ldots,\vec{u}\left(T\right)\right]$
as the matrix containing input vectors $\vec{u}(t)$ as its columns.
The element $U_{at}$ is the $a$th component of $\vec{u}(t)$. Similarly,
define the output matrix $X=\left[\vec{x}\left(1\right),\ldots,\vec{x}\left(T\right)\right]$
as containing output vectors $\vec{x}(t)$ as its columns. Define
the output-input correlation matrix is
\[
\frac{XU^{\top}}{T}=\frac{1}{T}\sum_{t=1}^{T}\vec{x}\left(t\right)\vec{u}\left(t\right)^{\top}
\]
Its $ia$ element is the time average of $x_{i}u_{a}$, or $\langle x_{i}u_{a}\rangle$.
Similarly, define the output-output correlation matrix
\[
\frac{XX^{\top}}{T}=\frac{1}{T}\sum_{t=1}^{T}\vec{x}\left(t\right)\vec{x}\left(t\right)^{\top}
\]
Its $ij$ element is the time average of $x_{i}x_{j}$, or $\langle x_{i}x_{j}\rangle$.
Note that ``correlation matrix'' is used to mean second moment matrix
rather than covariance matrix. In other words, the correlation matrix
does not involve subtraction of mean values. This is natural for sparse
nonnegative variables, but covariance matrices may be substituted
in other settings.
\begin{problem}
[Constrained optimization] \label{prob:Primal}Define the goal of
unsupervised learning as the constrained optimization

\begin{equation}
\max_{X\geq0}\Phi^{\ast}\left(\frac{XU^{\top}}{T}\right)\text{ subject to copositivity of }D-\frac{XX^{\top}}{T}\label{eq:ConstrainedOptimizationCopositivity}
\end{equation}
where $D$ is a fixed matrix and $\Phi^{\ast}$ is a scalar-valued
function that is assumed monotone nondecreasing as a function of every
element of its matrix-valued argument.
\end{problem}

Monotonicity is an important assumption because it allows us to interpret
the objective of Eq. (\ref{eq:ConstrainedOptimizationCopositivity})
as maximization of input-output correlations.

\subsection{Copositivity vs. nonnegativity}

\citet{seung2017correlation} introduced the principle
\begin{equation}
\max_{X\geq0}\Phi^{\ast}\left(\frac{XU^{\top}}{T}\right)\text{ subject to nonnegativity of }D-\frac{XX^{\top}}{T}\label{eq:ConstrainedOptimizationNonnegativity}
\end{equation}
which differs from Eq. (\ref{eq:ConstrainedOptimizationCopositivity})
only by the substitution of ``nonnegativity'' for ``copositivity.''
(Here nonnegativity of a matrix is defined to mean nonnegativity of
all its elements.) While the formalisms here are valid for arbitrary
$D$, a convenient choice is to set diagonal elements of $D$ to be
$q^{2}$ and off-diagonal elements of $D$ to be $p^{2},$
\begin{equation}
D_{ij}=\begin{cases}
q^{2}, & i=j,\\
p^{2}, & i\neq j
\end{cases}\label{eq:DesiredCorrelations}
\end{equation}
If $p$ is much smaller than $q$, the nonnegativity constraint $\langle x_{i}x_{j}\rangle\leq D_{ij}$
in Eq. (\ref{eq:ConstrainedOptimizationNonnegativity}) amounts to
decorrelation. 

A symmetric matrix $S$ is said to be copositive when $\vec{v}^{\top}S\vec{v}\geq0$
for every nonnegative vector $\vec{v}\geq0$. This constraint is analogous
to positive semidefiniteness but is more complex because it cannot
be reduced to a single eigenvalue constraint. \citet{hahnloser2003permitted}
give sufficient and necessary conditions for copositivity involving
eigenvalues of submatrices. 

Nonnegativity of $S$ is a sufficient condition for copositivity of
$S$, but it is not a necessary condition. In particular, copositivity
of $S-XX^{\top}/T$ does not require nonnegativity, so a solution
of Problem \ref{prob:Primal} may have $\langle x_{i}x_{j}\rangle>D_{ij}$
for some $i$ and $j$.

A necessary condition for copositivity of $S$ is nonnegativity of
its \emph{diagonal elements}, since $\vec{e}_{i}^{\top}S\vec{e}_{i}<0$
if $S_{ii}<0$ where $\vec{e}_{1},\ldots,\vec{e}_{n}$ denotes the
standard basis for $\mathbb{R}^{n}$. In particular copositivity of
$S=D-XX^{\top}/T$ requires that $\langle x_{i}^{2}\rangle\leq D_{ii}$
for all $i$. These inequalities will be called ``power constraints,''
because they limit the power in the outputs.

If either of the diagonal elements $S_{ii}$ and $S_{jj}$ vanish,
then a necessary condition for copositivity is nonnegativity of the
off-diagonal element $S_{ij}$. Therefore $\langle x_{i}x_{j}\rangle$
may exceed $D_{ij}$ in a solution of Problem \ref{prob:Primal} only
if the power constraints for $i$ and $j$ are not saturated.

\subsection{Correlation game from Legendre-Lagrangian duality}

The copositivity constraint in Eq. (\ref{eq:ConstrainedOptimizationCopositivity})
can be enforced by introducing Lagrange multipliers $A$ and $\Lambda$,
\begin{equation}
\max_{X\geq0}\min_{A,\Lambda\geq0}\left\{ \Phi^{\ast}\left(\frac{XU^{\top}}{T}\right)+\frac{1}{2}\trace A\left(D-\frac{XX^{\top}}{T}\right)A^{T}+\frac{1}{2}\trace\Lambda\left(D-\frac{XX^{\top}}{T}\right)\right\} \label{eq:Primal}
\end{equation}
The Lagrange multiplier $A$ is a nonnegative $r\times n$ matrix.
The outer maximum must choose $X$ so that $D-XX^{\top}/T$ is copositive
because otherwise the minimum with respect to $A$ is $-\infty$.
The Lagrange multiplier $\Lambda=\diag\left\{ \lambda_{1},\ldots,\lambda_{n}\right\} $
is a nonnegative diagonal matrix. The outer maximum must choose $X$
so that the diagonal elements of $D-XX^{\top}/T$ are nonnegative
because otherwise the minimum with respect to $\Lambda$ is $-\infty$.

As mentioned above, copositivity of $D-XX^{\top}/T$ by itself already
implies that the diagonal elements are nonnegative. It follows that
the Lagrange multiplier $\Lambda$ is redundant for the primal problem,
though it does affect the dual problem. Similarly, adding extra rows
to the Lagrange multiplier $A$ does not change the primal problem.
For enforcing the copositivity constraint, it would be sufficient
for $A$ to be $1\times n$. However, making $r>1$ does affect the
dual problem.
\begin{problem}
[Game between cells and connections]Switching the order of min and
max in Eq. (\ref{prob:Primal}) yields the dual problem,
\end{problem}

\begin{equation}
\min_{A,\Lambda\geq0}\max_{X\geq0}\left\{ \Phi^{\ast}\left(\frac{XU^{\top}}{T}\right)+\frac{1}{2}\trace\left(D-\frac{XX^{\top}}{T}\right)\left(A^{\top}A+\Lambda\right)\right\} \label{eq:Dual}
\end{equation}
This is an upper bound for Eq. (\ref{eq:Primal}) by the minimax inequality.

At this point, it is convenient to define the objective function $\Phi^{\ast}$
as the convex conjugate (Legendre-Fenchel transform) of a function
$\Phi$,

\begin{align}
\Phi^{\ast}(C) & =\max_{W\geq0}\left\{ \sum_{ia}W_{ia}C_{ia}-\Phi\left(W\right)\right\} .\label{eq:ConvexConjugate}
\end{align}
The nonnegativity constraint on $W$ in Eq. (\ref{eq:ConvexConjugate})
guarantees that $\Phi^{\ast}(C)$ is monotone nondecreasing as a function
of every element of $C$. The function $\Phi$ can be interpreted
as a regularizer or prior for the weight matrix $W$.

With Legendre duality, a maximization with respect to $W$ is implicit
in Eq. (\ref{eq:ConvexConjugate}). Switching the order of $W$ and
$X$ maximizations yields the following equivalent problem.
\begin{problem}
[Game between connections]\label{prob:DualGame}The Lagrangian dual
of the constrained optimization in Problem \ref{prob:Primal} is
\begin{equation}
\min_{A,\Lambda\geq0}\max_{W\geq0}R(W,A^{\top}A+\Lambda)\label{eq:CorrelationGame}
\end{equation}
with payoff function defined by
\end{problem}

\begin{equation}
R(W,L)=\max_{X\geq0}\left\{ \frac{1}{T}\trace W^{\top}XU^{\top}-\Phi(W)+\frac{1}{2}\trace\left(D-\frac{XX^{\top}}{T}\right)L\right\} \label{eq:Payoff}
\end{equation}
The min-max problem can be interpreted as a zero-sum game between
$W$ on the one hand and $A$ and $\Lambda$ on the other.

Problem \ref{prob:DualGame} is closely related to the correlation
game previously introduced by \citet{seung2017correlation},

\begin{equation}
\min_{L\geq0}\max_{W\geq0}R(W,L)\label{eq:CorrelationGameFullRank}
\end{equation}
Problem \ref{prob:DualGame} constrains $L=A^{\top}A+\Lambda$ for
some nonnegative $A$ and $\Lambda$, so it is an upper bound for
Eq. (\ref{eq:CorrelationGameFullRank}). This is the mathematical
interpretation of choosing a parametrized form $A^{\top}A+\Lambda$
for the Lagrange multiplier $L$, as was done by \citet{seung2018unsupervised}.

The network model of Section \ref{sec:NetworkModel} follows by setting
\[
\Phi(W)=\frac{\gamma}{2}\sum_{ia}W_{ia}^{2}+\frac{\kappa}{2}\sum_{i}\left(\sum_{a}W_{ia}\right)^{2}
\]
in Eq. (\ref{eq:Payoff}) and applying online projected gradient ascent
to perform the maximizations in Eq. (\ref{eq:CorrelationGame}) and
online projected gradient descent to perform the minimizations. For
a more general choice of $\Phi(W)$, Eq. (\ref{eq:UpdateW}) should
be replaced by

\[
\Delta W_{ia}\propto x_{i}u_{a}-\frac{\partial\Phi}{\partial W_{ia}}
\]

\section{Correlation game between cells}

The second normative principle concerns transformation of a sequence
of nonnegative input vectors $\vec{u}(1),\ldots,\vec{u}(T)$ into
two sequences of nonnegative output vectors $\vec{x}(1),\hdots,\vec{x}(T)$
and $\vec{y}(1),\hdots,\vec{y}(T)$. Define the input matrix $U=\left[\vec{u}(1),\ldots,\vec{u}\left(T\right)\right]$
and the two output matrices $X=\left[\vec{x}\left(1\right),\ldots,\vec{x}\left(T\right)\right]$
and $Y=\left[\vec{y}\left(1\right),\ldots,\vec{y}\left(T\right)\right]$.
\begin{problem}
[Game between cells] \label{prob:SecondGame}Define the goal of unsupervised
learning as the zero-sum game between $X$ and $Y$

\begin{equation}
\max_{X\geq0}\min_{Y\geq0}\left\{ \Phi^{\ast}\left(\frac{XU^{\top}}{T}\right)-\Psi^{\ast}\left(\frac{YX^{\top}}{T}\right)+\frac{1}{2}\trace\frac{YY^{\top}}{T}\right\} \label{eq:XYgame}
\end{equation}
where $\Phi^{\ast}$ and $\Psi^{\ast}$ are scalar-valued functions
assumed monotone nondecreasing as a function of every element of their
matrix-valued arguments.
\end{problem}

Note that only nonnegativity constraints remain in Problem \ref{prob:SecondGame};
the copositivity constraint of Problem \ref{prob:Primal} is completely
hidden. This correlation game can be interpreted as follows. The $E$
cells would like to maximize $S-E$ correlations (make $\Phi^{\ast}$
large) and minimize $E-I$ correlations (make $\Psi^{\ast}$ small).
The $I$ cells would like to maximize $I-E$ correlations (make $\Psi^{\ast}$
large) and minimize power (make $\trace YY^{\top}/T$ small). There
is conflict between the $E$ and $I$ cells because $E$ cells would
like to minimize $E-I$ correlations while $I$ cells would like to
maximize them. The compromise is that $E$ cells incompletely decorrelate
from each other.

Problem \ref{prob:SecondGame} is equivalent to Problem \ref{prob:Primal}
given the definition of

\begin{equation}
\Psi^{*}(C)=\max_{A\geq0}\left\{ \sum_{i\alpha}A_{\alpha i}C_{\alpha i}-\Psi(A)\right\} \label{eq:ConjugatePriorA}
\end{equation}
as the Legendre transform of
\begin{align}
\Psi(A) & =\frac{1}{2}\trace ADA^{\top}\label{eq:PriorA}
\end{align}

\begin{proof}
Substituting the definitions of Eqs. (\ref{eq:ConjugatePriorA}) and
(\ref{eq:PriorA}) into Eq. (\ref{eq:XYgame}) yields

\begin{align*}
\max_{X\geq0}\min_{A,Y\geq0}\left\{ \Phi^{\ast}\left(\frac{XU^{\top}}{T}\right)-\frac{1}{T}\trace A^{\top}YX^{\top}+\frac{1}{2}\trace ADA^{\top}+\frac{1}{2}\trace\frac{YY^{\top}}{T}\right\} 
\end{align*}
This is minimized when $Y=AX$, attaining the value
\begin{align*}
\max_{X\geq0}\min_{A\geq0}\left\{ \Phi^{\ast}\left(\frac{XU^{\top}}{T}\right)-\frac{1}{2T}\trace A^{\top}AXX^{\top}+\frac{1}{2}\trace ADA^{\top}\right\} 
\end{align*}
This is identical to Eq. (\ref{eq:Primal}), except for the omission
of the Lagrange multiplier $\Lambda$ which, as mentioned previously,
is redundant in the primal problem.
\end{proof}

\section{Discussion}

The second normative principle is also interesting because it can
be generalized to include $E-E$ and $I-I$ connections. This will
be the subject of future work.

\section*{Acknowledgments}

The author is grateful for helpful discussions with J. Zung, C. Pehlevan
and D. Chklovskii. The research was supported in part by the Intelligence
Advanced Research Projects Activity (IARPA) via DoI/IBC contract number
D16PC0005, and by the National Institutes of Health via U19 NS104648
and U01 NS090562.

\bibliographystyle{plainnat}
\bibliography{/home/sseung/HebbAnti/doc/Foldiak}

\end{document}